\newcommand{\beq}{\begin{quote}}
\newcommand{\enq}{\end{quote}}
\newcommand{\be}{\begin{equation}}
\newcommand{\en}{\end{equation}}

\newcommand{\om}{\omega}
\newcommand{\Om}{\Omega}

\documentclass[preprint2]{aastex}
\begin{document}
\title{ Determination of masses and other properties of extra-solar planetary
systems with more than one planet}
\author{Michael Nauenberg}
\affil{Physics Department, University of California,
    Santa Cruz, CA 95064}
\begin{abstract}
Recent analysis of the  Doppler shift oscillations of the light 
from  extra-solar planetary systems 
indicate that some of these systems have  more than one  
large planet. In this case it has been shown that
the  masses of these planets can be determined 
without the familiar ambiguity due to the unknown inclination 
angle of the plane of the  orbit of the central star provided, 
however, that its mass is known.  A method  is presented here
which  determines  also  a lower limit to the mass of 
this  star from these observations. As an illustration, our 
method is applied to the Keck and Lick data for GJ876.

\end{abstract}

\section{Introduction}

During the past six years  a large number of extra solar planetary systems have 
been discovered by observations of Doppler shifted oscillations
of the light emitted by the central star \citep{udry}. Recent  
analysis of the data has shown that some of these systems  have  more
than one Jupiter-size  planet circulating the central star
\citep{marcy2}.  It is well known that when there is 
only a single planet an ambiguity occurs  in the
determination of the  mass of this  planet when
the inclination angle of the plane of the orbit is not known.
For systems with more than one  sizable  planet, however, 
it turns out that this ambiguity can be removed when 
the gravitational interaction  between these planets is
important to the  evolution of the planetary 
orbits, as has been pointed out by Laughlin and Chambers (2001).  
By the  equivalence principle  this interaction 
is  proportional to the mass of the planets, thus
providing   an  additional   dependence on these masses 
which  is absent  when there is only a single planet. 
In the case of multiple planets
only approximate  analytic solutions of the gravitational
equations of motion exist, and one must resort to  numerical integrations
to analyze the data.
In this paper we present a method based on such an integration
to obtain the masses and other  properties of the extra-solar
planetary system. As an illustration we apply our method  to
the Keck and Lick data  for GJ876 obtained by Marcy et. al. (2001).
Some previous  analyses of this data depended  on the approximation
that each of these planets is traveling on a
Keplerian elliptical orbit with either constant orbital elements, Marcy et. al.
(2001),  or on  variable orbital  elements \footnote{ When there is  more
than one planet, such an approximation is not unique. For example,
one can locate the center of attraction and corresponding
focus of each of  the ellipses either at the
center of mass or at the position of the star leading to somewhat different
values for the orbital elements.}, Laughlin and Chambers (2001),
although these latter authors have implemented also an exact numerical
integration of the equations of motion . While we have been able to  
verify  their results with our method, we have found a second solution 
which differs significantly from theirs. The value of the  reduced 
chi-square of these two fits is of order 2.5 - 3, although
the mean is approximately one, indicating  that some 
fundamental physics in the analysis of the data has not been taken into 
account. Indeed, an additional  source for velocity fluctuations in the light
emitted by the central star may  be  due to the convective motion 
and turbulence of the chromosphere, which has been shown to be correlated to
increase magnetic activity in some stars, Saar, Butler and Marcy (1998), 
Saar and Fischer (2000).
An estimate of the magnitude of these fluctuations  can  be obtained 
from the requirement that the reduced chi-square be of order unity,
as will be shown in  section 3.

\section {Method for analysis of extra-solar planetary data }
Our starting point is to  re-scale  the gravitational  
equations of motion by introducing a length scale $l$ and time scale $\tau$ 
which satisfies   the Kepler-Newton relation
\be
\label{kepler}
\frac{l^3}{\tau^2}=Gm_s,
\en
where $m_s$ is the mass of the central star. This mass 
is taken as one of the  parameters in a least-square fit to the data,
while either $l$ or $\tau$ can be chosen as another parameter. 
In these rescaled variables the magnitude of the force per unit mass  due
to a  planet with mass $m^g_j$  at a distance $r_j$ is  
is   
\be
\label{inter}
(m^g_j/m_s)(1/r_j^2)
\en
For clarity  we have labeled this  planetary  mass
with a superscript $g$ to indicate
that we are referring here to the  {\it gravitational} masses.  
On the other hand, by momentum conservation, the velocity of the central star 
is related to the velocities
$v_j$ of $n$  planets according to the  relation
\be
\label{vel1}
\vec v_s=-\sum_{j=1}^{j=n} \frac{m^i_j}{m_s}\vec v_j +\vec v_o 
\en
which depends on  the {\it inertial} masses $m^i_j$ of the planets
which we label  by the superscript $i$.
Here, the velocity $v_o$ is proportional to the uniform velocity 
of the center of mass relative to the observer, and becomes 
another  parameter in our  fit.
By the equivalence principle, the gravitational
and inertial masses are equal, $m^g_j=m^i_j=m_j$, which implies that 
the same  mass ratio $m_j/m_s$ appears 
in both this kinematical relation and  in the 
dynamical interaction between planets, Eq. \ref {inter}. This is the key 
which opens  a new way to obtain these  mass ratios
in extra-solar planetary systems with  more than one planet.
\footnote {In the future, with increasing data, 
these  two mass ratios could be take as independent
parameters in a fit  to  provide a new  test of the equivalence principle 
for extra-solar planetary systems}
According to our  scaling, Eq. \ref {kepler}, the  observed velocities 
are  obtained by multiplying these  velocities 
by  a scale factor
\be
\label{vscale}
V= \frac{l}{\tau}=(\frac{ G m_s}{\tau})^{(1/3)}
\en
The observed Doppler variations, however, depend also  on the inclination angle
$i$ of the mean plane of the orbit of the star relative to
the observer, and  therefore the relevant scale for observations
is the product   $Vsin(i)$. Consequently, the independent parameters in
a fit to the observations  are the product  $m_s sin^3 (i)$  
and the mass ratios $m_j/m_s$.

The additional  parameters which are  required  to determine  
the  evolution of the extra solar system
are those parameters  which determine the  initial conditions, 
e.g. the  position and velocities of each of the planets at the
start of the observations, and the relative uniform velocity of the
center of mass of the system. The initial velocity of the
central star is then given by the conservation of momentum relation
Eq. \ref {vel1}. A useful  approach, currently
in practice, is to parameterize  the  orbital elements
of the initial osculating Keplerian  ellipse for each planet.
It must be remembered, however, that when interplanetary
perturbations are important  these orbital elements evolve in
time. As will be shown below, some of these elements  
have  periodic variations, 
while  others change continuously. Hence, in general these initial parameters 
do not  describe the  mean properties of the system
when interplanetary perturbations are important.

The  values of the  parameters are obtained by a least
square fit to the data  obtained by integrating numerically
the gravitational equations of motions. 
In particular, we consider  the dependence of this fit
as a function of our  parameter $m_s sin^3(i)$. According to Eqs. 
 \ref {vel1} and  \ref {vscale}, 
when  this parameter is increased above the value at which
a minimum  has been found, 
the  planetary mass ratios $m_j/m_s$ 
are expected to  decrease inversely with the cube
root of  $m_s sin^3$, in order to fit the observed
velocity of the star. Hence, if the inter-planetary interactions are important
the chi-square of the fit  should increase;  otherwise it approaches a constant.
Correspondingly, when this parameter is  decreased 
the chi-square also  increases, because now the interplanetary 
interactions increases above their actual value.
Since $sin(i)\le 1$, the value of the parameter 
$m_s sin^3(i)$ at the minimum chi-square  
then gives a {\it lower limit} for the  mass $m_s$.
When this mass can also be obtained from the mass-luminosity 
relation,  our method   provides a novel check on
the validity of this relation as well as  on the
self-consistency of the least-square fit.

\section {Application to GJ 876}

We illustrate our method by an application
to the  Keck and Lick data for the velocity modulations of GJ876  
obtained by Marcy et al (2001).  
These authors have  fitted  their data assuming that there are 
two large planets orbiting the center of mass on Keplerian ellipses
which have  {\it constant} orbital elements, 
and they found that the periods were  nearly in a  2:1  ratio.
The occurrence of this resonance, however, indicates
that the interactions between these planets cannot be neglected
as was assumed originally. 
Recently, Laughlin and Chambers (2001) also fitted  this data by integrating
the equations of motion numerically, but we have found that their solution
is not unique, and we present here  another fit which gives 
planetary masses and other properties of the system which differ significantly
from their results.

In Fig. 1  we  show the dependence of the reduced  chi-square
\footnote{We use here the conventional definition of  reduced chi-square  
which corresponds to
the square of the quantity called ``reduced chi-square" in  papers
on GJ876 listed in the references.} 
obtained by a least-square fit to the Keck data, as a 
function of the parameter $m_s sin^3(i)$. 
The minimum in the parameter space  was found by a simplex program.
Under the assumption that the two planets,
and consequently the central star, all move on the same plane, 
we have eleven  parameters in our fit. These  correspond to the 
two planetary masses, the eight parameters which determine the  
initial positions and velocities of the planets 
in a plane at the time of the first Keck observation, and the component of 
relative uniform  velocity of the center of mass of the
system, $v_o$  along the line of sight.
Alternatively, these  parameters can be chosen to be  the initial
orbital elements of the osculating ellipses for the 
inner and outer planets. These  elements are given in Table 1  
with $v_o$=77.7 m/s.  

We find  that the  minimum of the chi-square lies
at $m_s sin^3(i) \approx .3$ solar masses which 
is  remarkably close to the value of the mass of the central
star  obtained by Marcy et. al. (2001)
from the mass-luminosity relation, $m_s=.32(.05)$ solar masses.
The value of our reduced chi-square
square  of about 3 at the minimum, however, would seem 
to indicate  a very poor 
fit to the Keck data. If the  observational errors have 
not been underestimated, this must be due 
to physical sources for fluctuations of the Doppler 
shifted light  which have
not been taken into account yet. It is known that
velocity fluctuations can  occur due to the convective 
motion and turbulence  in the chromosphere of a star,   
Saar, Butler and Marcy (1998), Saar and Fischer (2000).
Therefore, the  mean square of these  fluctuation  should be added 
to the square of the observational errors to obtain
the actual reduced chi-square. Assuming that the reduced
chi-square for our fit should be of order unity, we 
obtain an estimate for the magnitude of these fluctuation 
of 4-6 m/s comparable 
to the observational errors in the Keck
data, which is 3-5 m/s. These fluctuations  may also limit 
the accuracy to which the  gravitational effects due to planets on the
motion of the star can be observed.

An independent  confirmation of our fit is  obtained
by applying the parameters from the Keck data 
to evaluate the reduced chi-square for the Lick data points.
This fit,  shown in Fig 2., has only a single parameter corresponding
to the  relative  zero- point velocity  between the Keck and Lick 
telescope systems. It can be seen that  the reduced chi-square 
also increase rapidly for $m_s sin^3(i)$  below $.3$, but above this
value it now  approaches a constant.  Since
the statistical errors in the Lick data 
are about 3 times larger than those in the Keck data, evidently
this fit is not very  sensitive
to interplanetary perturbations of the order of, or smaller than,
its actual value. 
For  $m_s=.32$ solar masses, these  results imply  that $sin(i) \approx 1$. 
In  contrast,  Laughlin and Chambers (2001) found
that  $sin(i)=.55$ from a fit to the Keck data, 
and $sin(i)=.78$ for a corresponding fit to the combined Keck and Lick data. 

In Fig. 3 we show the dependence of the resulting  masses of the two
planets as a function of $m_s sin^3(i)$, where we have chosen
for the planetary  mass scale the observed magnitude of the central star.
This dependence is well fitted by the  relation 

\be
m_j=m_{jo} (.32/(m_s sin^3(i))^{(1/3)}
\en
where $m_{jo}$ is a constant which is normalized  at $m_s(sin(i)=.32$.
From the  minimum of our chi-square, we obtain 
$m_1=.6$ and $m_2=1.9$ in units 
of  Jupiter's mass, which is comparable to the results of  Marcy et. al. {2001},
but in disagreement  with two different  values for these 
planetary masses obtained by  Laughlin and Chambers (2001)
who found substantially larger values due to their
smaller values of sin(i).

In Fig. 4 we show our result for the velocity modulations  
of the central star, with the Keck and Lick data points
superimposed as squares and pentagons respectively.
A blow-up of this plot  is  shown in Fig.  5 which 
exhibits the characteristic mid period oscillations which
are signatures of the inner planet.
The zero point in the time scale  has been chosen 
at the first Keck data point, and we have 
extended our calculation to 4000 days to show the
occurrence of a long term periodic modulation of about
3200 days of the rapid  oscillations of about 60 days .  
These rapid oscillations are associated with the mean period
of the outer planet, while the long term modulation
is associated with the  nearly uniform  rotation
of the  major axis of the orbit of  inner planet, as will
be demonstrated in the next section.
The apparent symmetry of these oscillation  on reflection
about an axis at $ v_o\approx 78 m/s$, Eq. \ref {vel1},
shifted by half the long modulation period, 
is due to the fact that  this major axis rotates 
through $180^0$ during  this half  period. In addition, we see that
the envelop exhibits also a  medium length modulation of about
660 days which, as  we shall see, correspond to the mean period 
of oscillations of the  major axis of the planets and the eccentricity
of the inner planet, and is associated with the oscillations  from
an exact 2:1 resonance which will be  discussed  later on.

\section {Properties of the two  planetary orbits of GJ876}

From the numerical solution of the equations of motion  
one can determine directly the properties of the planetary orbits 
and the orbit of the central star.  A typical example of the
planetary orbits  during a
a single period of the outer planet is shown in Fig. 6.  During this
time interval, the inner planet turns approximately twice around 
the central star traveling along two  orbits 
which are slightly displaced relative
to each other due to the interplanetary perturbations
and the motion of the central star. At the time that the inner
planet first reaches its  pericenter (triangle) the outer planet 
(triangle)  is nearly aligned with the central star 
which is shown on this scale  only as a dot. Approximately half a period later 
the inner planet has completed a revolution,  and it is again at 
pericenter (square)  while the outer planet (square) comes  again 
into conjunction with the central star, but at the opposite side of the 
initial location on its orbit. Then, after the inner planet completes
a second revolution, the outer planet also returns to conjunction with
the star. This is the characteristic signature of a dynamical
2:1 resonance.  The resulting
motion of the central star with its  position
at ten equal time interval  is shown in Fig. 7. When the inner and
outer planet are aligned  on the same side of the star we see
that the motion of the star is accelerated, while when these 
planets are in conjunction on opposite sides of the
star  the motion is slowed down and a
dimple appears in the orbit. These planets exchange angular momentum
in an oscillatory fashion as shown in Fig. 8.
The orbital periods of the planets are obtained by
evaluating the time elapsed between successive passages at nearest distance, 
$r_p$, or largest distance, $r_a$, from the center of mass of the extra-solar 
planetary system. While the ratio of the
mean periods of the outer  and inner planet is approximately  two,
as obtained  previously by Marcy et.al (2001), we find that 
these periods are not constants as had been  assumed previously. 
As we have seen, during a  single period of the outer planet
the inner planet travels around two slightly different orbits
with periods which each have a periodic oscillation
of 660 days as shown in Fig. 9. 
In contrast, the variations of the period of the outer planet
shown in Fig. 10, exhibits  a longer term periodicity of
3200 days associated with the rotation period of the axis
of the orbit of the inner planet. This period differs somewhat
when it is defined relative to $r_p$ or $r_a$.

Correspondingly, we can define also an effective  major 
axis $ a=(1/2)(r_a+r_p)$
and eccentricity $e=(r_a-r_p)/(r_a+r_p)$
for each planetary period. The results for the 
inner planet are shown  in  Fig. 11 and 12  which exhibit
again  periodicity around  two  orbits. 
While the major axis  of  the outer planet exhibits
similar oscillations with a period of 660 days, Fig. 13, its 
eccentricity has  a much  longer periodicity 
of about 3200 days, Fig. 14, associated with
the rotation period of the major-axis of the inner planet, 
which give rise to  the long term modulation period  
shown in Fig. 4. This is shown 
in Fig.15 where  we see  that, apart from  small 660 day oscillations,
the longitude of the  inner planet at pericenter rotates with nearly uniform
angular velocity, and it is always approximately  
aligned at that time with the longitude of the
central star and the outer planet, in accordance with a dynamical
2:1 resonance. In Fig. 16 we show the corresponding rotation
of the pericenter of the outer planet which exhibits rapid changes
when the eccentricity decreases rapidly, see Fig. 14.   

\section {The 2:1 resonance in GJ876}

The occurrence of a  2:1 resonance in GJ876   keeps the inner and outer 
planets from getting too close to each other which can  cause large  
perturbations which would disrupt the system. 
In the present case, when the eccentricity of the outer planet
is very small compared to that of the inner planet, 
we require that 
each time  the inner planet completes two turns around the 
center of mass of the system and returns to 
pericenter, the outer planet turns around only once, and becomes then 
aligned with the inner planet, the central star, 
and the center of mass of the system.
That this condition is in fact  approximately satisfied 
by GJ 876 can be seen in Fig. 6  which shows  typical orbits  
of the inner and outer
planet for two revolutions of the inner planet and one revolution
of the outer planet. Analytically,
this  condition for a 2:1 resonance can be written  in the  form
\be
\int_0^{2P_i}\om_o(t)dt=2\pi-\int_0^{2P_i}(\Om_i(t)-\Om_o(t))dt
\en
where $\om_o(t)$ is the angular velocity of the outer planet,
$\Om_i(t)$ and $\Om_o(t)$ are the  rotation rates of the
major axis of the inner and outer planets described previously, 
and $P_i$ is the mean period of two successive rotations of the inner planet.
We assume here that the direction  of rotation of the two  planets is the same,
but  in opposite direction to that of the major axis.  
For  $2P_i\approx P_o$  we obtain
\be
\label{res1}
(P_o-2P_i)\om_o(P_o)= \int_0^{2P_i}(\Om_i(t)-\Om_o(t))dt
\en
From this expression one can obtain  the conventional relation
for a 2:1 resonance,  Murray and  Dermott (2001),
by replacing the rotation rates
by their mean values, and by  assuming also  that 
the rotation rate of the outer planet is uniform or that 
$w_o(P_o)=2\pi/P_o$,
This latter condition, however,  is not valid in  general.
The deviation from exact resonance gives rise to  librations
which  have a mean  period of 660 days,
while  the  rotation of the axis of the orbit of the  planets 
has a mean  period of  3200 days.
These librations are the origin of the oscillations
in the orbital elements of the planetary orbits shown
in Figs. 9-11 and 12-13,  and  can also  be seen directly  from the
the data by observing the oscillations in the
boundaries of the the fit to the Keck data,  shown in Fig. 4.
Sinusoidal librations  were  introduced  by Laughlin and Chambers (2001)
in an analytic model to a  fit to  the Keck and the Lick  data
which included oscillations in the major axis of the orbit planets,
but their model  neglected the  corresponding  oscillations in the eccentricities
and  periods, and the occurrence of a rapid oscillation with mean  period  
$P_i$ between two effective  
elliptic orbits which characterize the motion of of the inner planet. 
The identification of two long term periodicities
indicates that the 2:1 resonance motion in GJ876  
is quasi-periodic.

\section {Concluding remarks}

The  essential new  feature in our analysis is the scale transformation,
Eqs. \ref{kepler} and  \ref{vscale}, which shows that the mass $m_s$ of the
central star and the inclination angle $i$ of the mean plane of its  orbit
are not independent parameters in a least square fit to the
observational data. Instead, these two parameters appear as a single variable
in the form of a product
$m_s sin^3(i)$ which provides a lower limit to the mass
of the star directly from observations of Doppler shifted oscillations
of the emitted light.
We have  emphasized that the scaled gravitational interactions depends only 
on the ratios $m_j/m_s$ 
of the planetary masses to the mass of the central
star, which is the reason why  these ratios can be determined independently of
a knowledge of the inclination angle $i$.
When the effects of interplanetary interactions are important,
the initial orbital elements of the osculating ellipses
for the planetary orbits, which are commonly introduced in the analysis of
extra solar planetary systems, do not correspond to mean properties 
of these systems.
We  have shown that for a 2:1 resonance, the motion of the inner planet
is actually characterized by a continuous
switching back and forth between two elliptical orbits during a single 
period of the outer planet as can be seen in
Figs. 8 to 11. For the case of GJ876  the eccentricity for the outer 
planet oscillates between a minimum value
of .006 and a maximum of .034, Fig. 14,  while its initial value
is .0018. The orientation of the major axis is not fixed,
but rotates nearly uniformly in the case of the inner planet
with a period of 3200 days, see  Fig. 15,
while for the outer planet it exhibits  a somewhat 
more complicated motion, shown in  Fig. 16.
The  rapid variation which appear  here around  t=3000 day occur near the
minimum of the eccentricity of the outer plane, and is associated
with the degeneracy for the major  axis in the limit of a circular orbit.
It is important to check   the long time stability
of any  numerical solution of the equations of motion, 
as has been pointed out by Laughlin and Chambers(2001), and this has been
verified  for the solution presented here (Laughlin, private
communication).
In our analysis we neglected the difference between
the inclination angles
of the mean planes of the planetary orbits,
and possible effects  due to tidal distortions of the planets.
The differences  between our fit to the data for GJ876  and
the corresponding fit of Laughlin and Chambers(2001)
indicates that at present it is not yet  possible
to determine  uniquely the properties of this system,
but we expect that this  ambiguity will be resolved by
additional data from future observations.
 
{\it Note added}.- After the completion of this work,  Marcy et. al (private
 communication)  released  9 new data points, and revised the  values
 for the velocities of the central star published previously.
 Our  fit to  this  data gives a similar reduced chi-square as before, 
 and the result showing the  last 9 data  points is given in  Fig. 17.
 In addition,  I  have  been informed
 that  two additional papers  analyzing
 GJ876 have appeared recently, one by Rivera and Lissauer (2001), 
 and the other by Lee and Peale (2001).

\subsection*{Acknowledgments}

I would like to  thank John Chambers, Don Coyne, Greg Laughlin, Deborah Fischer, 
Geoff Marcy,  Stan Peale, and  Steve Vogt for helpful comments, 
and the Rockefeller Foundation
for their hospitality at Villa Serbelloni in Bellagio, Italy,
where this work was completed.

\email{michael@mike.ucsc.edu}.

{}

\begin{figure}
\plotone{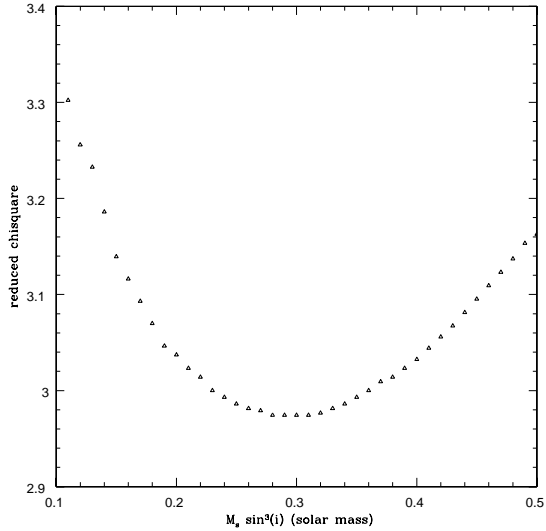}
\caption{
An eleven parameter least-square fit to the Keck data as
a function of the parameter $m_s sin^3(i)$
}
\end{figure}

\begin{figure}
\plotone{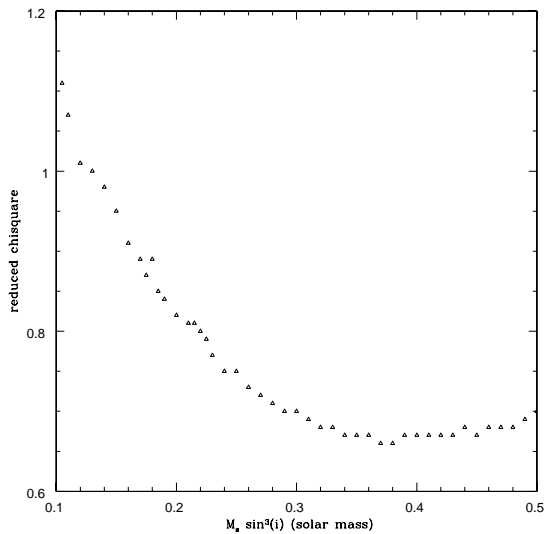}
\caption{
A single parameter least-square  fit to the Lick data
based on the best fit to the Keck data
}
\end{figure}

\begin{figure}
\plotone{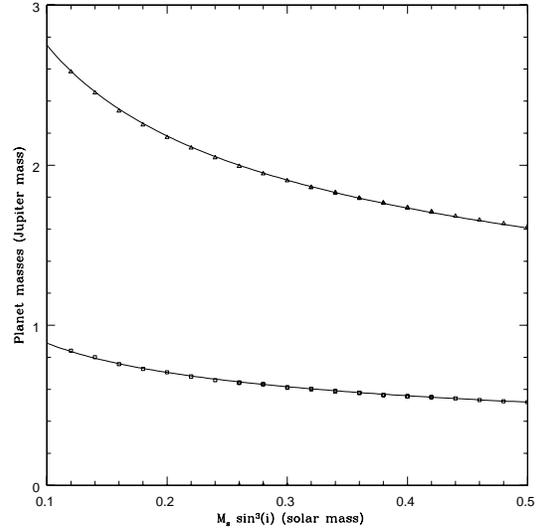}
\caption{
Masses of the inner and outer planet as a function
of the parameter $m_s sin^3(i)$
}
\end{figure}

\begin{figure}
\plotone{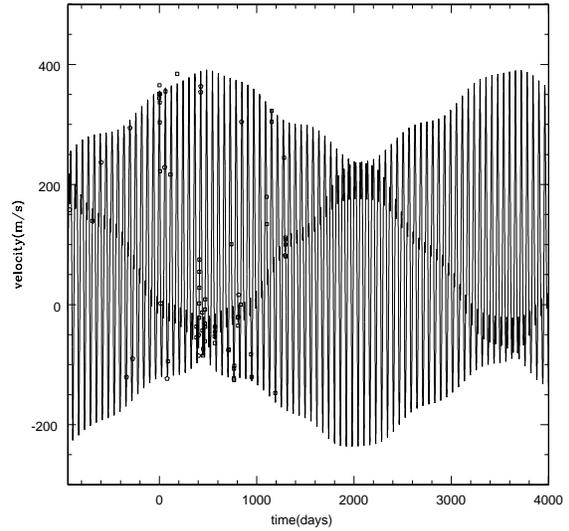}
\caption{
Calculated velocity oscillations of the
central star in GJ 876. The  Lick and Keck data
is shown as pentagons and squares respectively
}
\end{figure}

\begin{figure}
\plotone{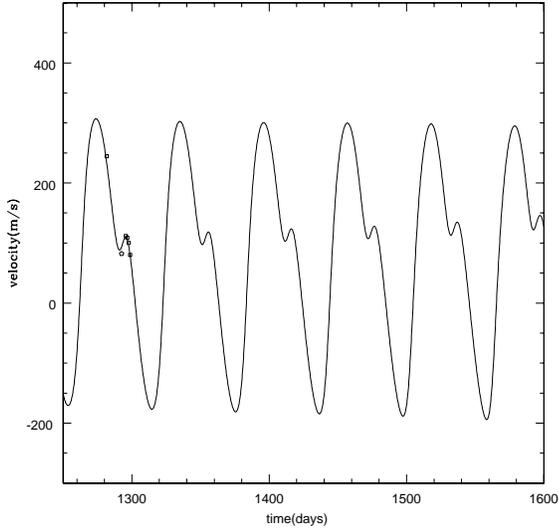}
\caption{
A blow up of Fig. 4 for the  velocity oscillations of the
central star in GJ 876 which exhibits the variations
at half period  characteristic  of the 
inner planet. 
}
\end{figure}

\begin{figure}
\plotone{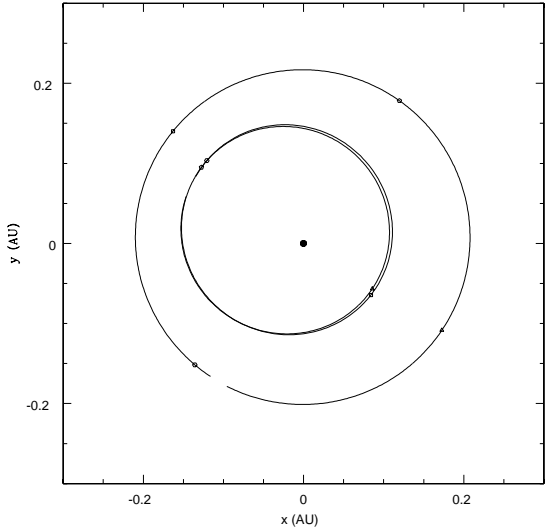}
\caption{
Orbits of the outer and inner planet showing the
resonance alignment of these planets at resonance with
the central star, triangles and squares, and their location 
when he outer planet is at quadrature, pentagons.
}
\end{figure}

\begin{figure}
\plotone{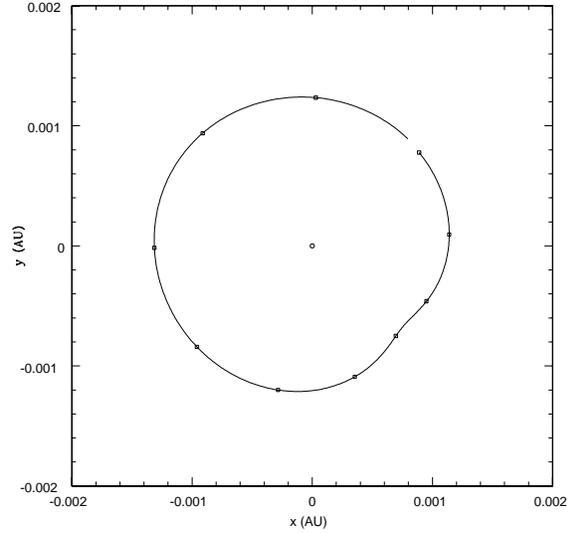}
\caption{
Orbits of the central star  showing its
position at ten equal time intervals
during a single period of the outer planet.
}
\end{figure}

\begin{figure}
\plotone{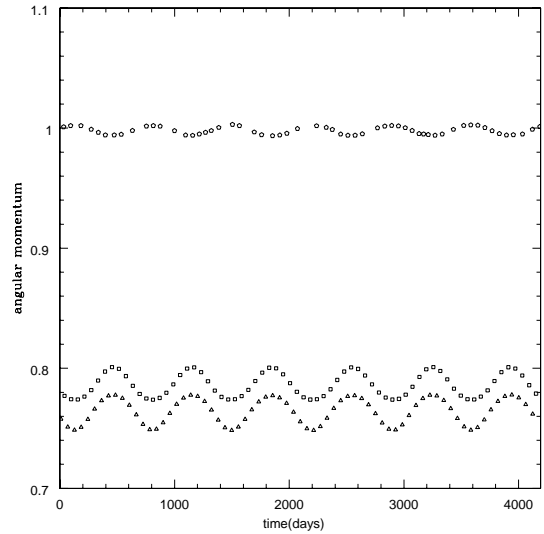}
\caption{
Angular momentum of the inner planet,
triangles and squares, and outer 
planet, pentagons,  at pericenter.
}
\end{figure}

\begin{figure}
\plotone{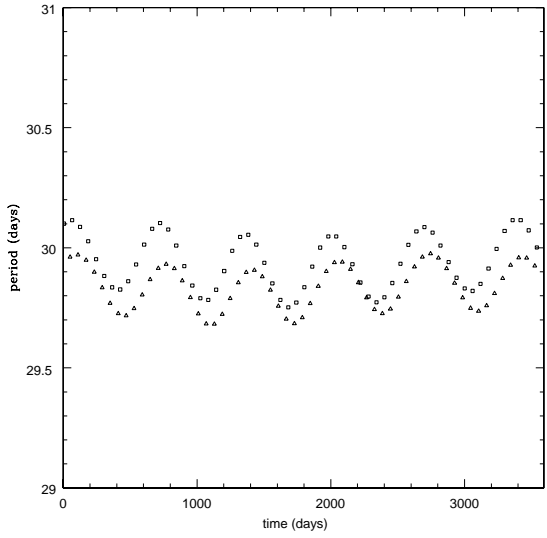}
\caption{
Period oscillations  for two successive 
orbits of the inner planet.
}
\end{figure}

\begin{figure}
\plotone{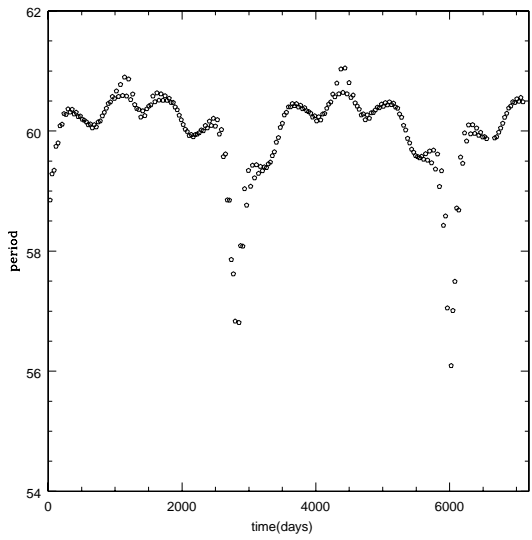}
\caption{
Oscillations for the period  of the outer planet.
}
\end{figure}

\begin{figure}
\plotone{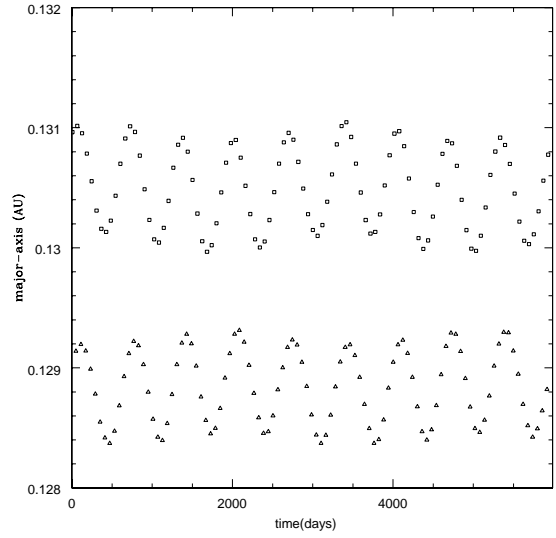}
\caption{
Major axis oscillations of two successive  orbits 
of the inner planet.
}
\end{figure}

\begin{figure}
\plotone{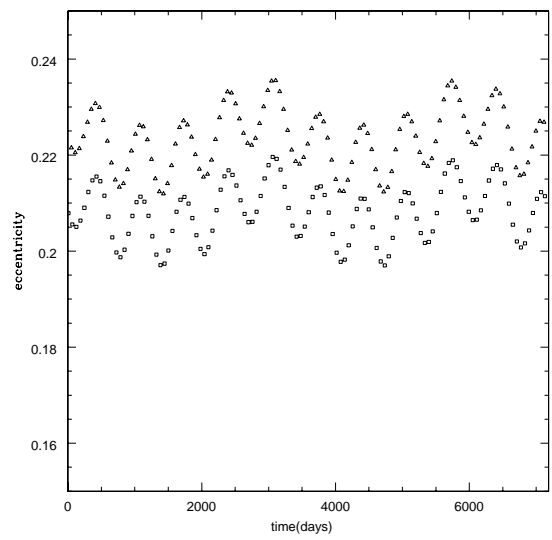}
\caption{
Eccentricity oscillations of  two successive orbits
of the inner planet. 
}
\end{figure}

\begin{figure}
\plotone{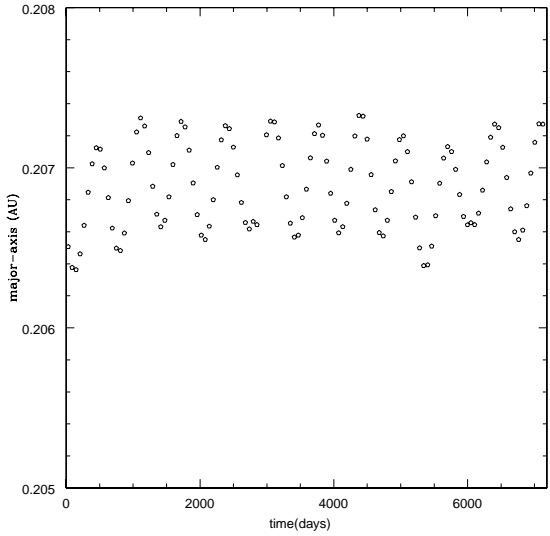}
\caption{
Major-axis oscillations of the outer planet.
}
\end{figure}

\begin{figure}
\plotone{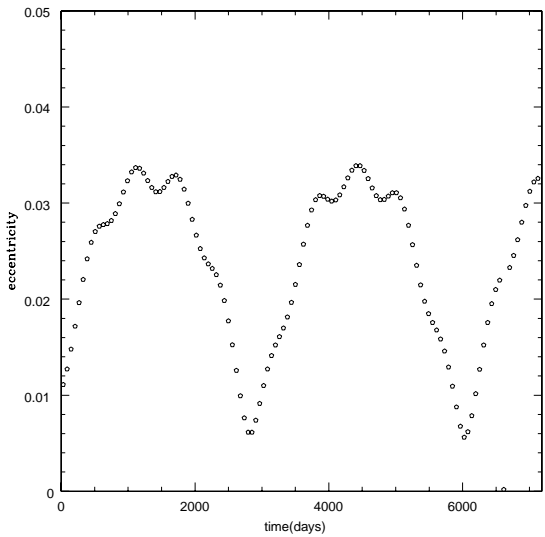}
\caption{
Eccentricity variations of the outer planet.
}
\end{figure}

\begin{figure}
\plotone{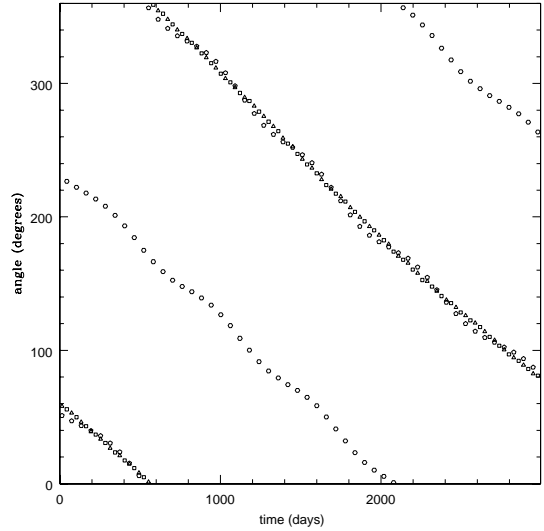}
\caption{
The longitude of the  inner planet at pericenter
for successive orbits (triangles and squares),  
and the  corresponding longitudes  of the outer
planet ( pentagons and hexagons). 
}
\end{figure}

\begin{figure}
\plotone{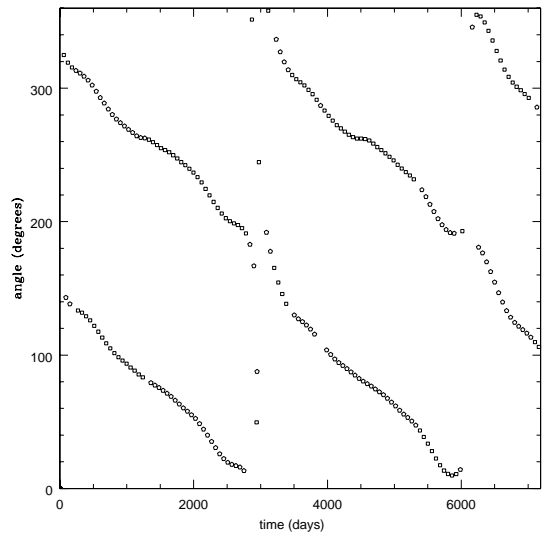}
\caption{
The longitude at maximum  and minimum distance
from the center for the outer planet  
}
\end{figure}

\begin{figure}
\plotone{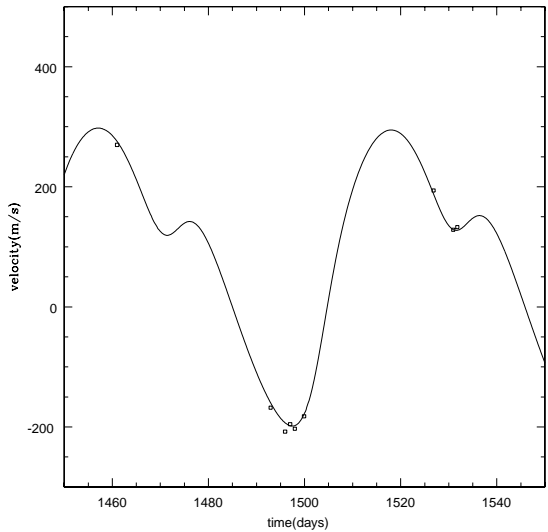}
\caption{
Least square fit to latest  Keck data point (squares).
}
\end{figure}

\begin{table}
\begin{center}
\begin{tabular}{lcc}
\hline
\hline
\\ Parameter & Inner & Outer\\
& \multicolumn{2}{c}\
\\
\hline
\\
Period\ (days)   &29.24     &59.90 \\
Mass ratios ($M_J$/$M_S$)      &.00180     &.00557\\
Major axis (AU) &.127      &.205 \\
Eccentricity    &.228      &.00185\\
pericenter (degrees) &-60.88    &29.39\\
ecc. anomaly     &269.83       &43.85\\
\\
\hline
\end{tabular}
\end{center}
\caption{\bf Initial osculating orbital elements and planetary  mass ratios for 
$m_s sin^3(i)=.32 $ solar masses and $v_o=77.7 m/s$} 
\end{table}

\end{document}